\documentclass[prd,amsmath,amssymb]{revtex4}

\usepackage{graphicx}
\usepackage{bm}

\normalbaselines \topmargin -0.25 truein \oddsidemargin 0.30
truein \evensidemargin 0.30 truein 

\begin{document}
	
	\title{Axion electrodynamics on the Bianchi spacetime platform:  \\ Fingerprints of shear of the aether velocity}
	
	\author{Amir F. Shakirzyanov}
	\email{shamirf@mail.ru} \affiliation{Department of General
		Relativity and Gravitation, Institute of Physics, Kazan Federal University, Kremlevskaya
		str. 18, Kazan 420008, Russia}
	\author{Alexander B. Balakin}
	\email{Alexander.Balakin@kpfu.ru} \affiliation{Department of General
		Relativity and Gravitation, Institute of Physics, Kazan Federal University, Kremlevskaya
		str. 18, Kazan 420008, Russia}
	
	\date{\today}
	
	\begin{abstract}
		In the framework of axionically extended Einstein-Maxwell-aether theory we study the structure of the electromagnetic field allowed by the anisotropic cosmological spacetime platforms associated with the Bianchi models. These models guarantee that the aether velocity possesses the shear, and we focus on its role in the evolution of the axion-photon systems. In this short note we discuss the extended master equations obtained under the assumption that the square of the shear tensor is included in the potential of the axion field.
	\end{abstract}
	
	\maketitle
	
	\section{Introduction}\label{Intro}
	
	The Bianchi models, which describe evolution of the homogeneous anisotropic Universe, are well known and very fruitful in the context of modeling of the physical system dynamics. We study the properties of the axion-photon coupling mediated by the dynamic aether \cite{1,2,3,4}. When we deal with the anisotropic Universe, the velocity of the aether $U^j$ is characterized not only by the expansion scalar $\Theta=\nabla_kU^k$ (e.g., as in the case of Friedmann Universe), but also by the nonvanishing shear tensor $\sigma_{pq}$. Thus, the Bianchi spacetimes become the unique models for the extended analysis of the aether dynamics and its impact on the electrodynamic systems. However, the constraint for the ratio of the velocities of the gravitational and electromagnetic waves, which has been obtained due to the observation of the binary neutron star merger (see  \cite{170817}), crossed out the shear tensor from the basic scalar $K^{ijmn}\nabla_iU_m \nabla_jU_n$ introduced into the Lagrangian of the dynamic aether theory \cite{J}. This means that in the canonic theory of the dynamic aether the shear tensor is a hidden quantity, and we have to search for another way to reveal the shear fingerprints. In this short note we suggest to consider a specific model of interaction between the axion field $\phi$ and aether velocity, namely, we introduce the square of the shear tensor into the axion field potential $V$. Since the axion field is coupled to the electromagnetic field, the shear of the aether velocity affects the structure of magnetic and electric fields in the Universe. In other words, we suggest the mechanism of the degeneration removal with respect to the shear of the aether velocity, based on the axion-aether interaction. Keeping in mind this idea, below we derive the extended master equations for the axion, vector, electromagnetic and gravitational fields.

	\section{The formalism}
	
	We start with the action functional of the Einstein-Maxwell-axion-aether theory
	\begin{equation}
	S_{({\rm EAA)}} =  \int d^4 x \sqrt{{-}g} \ \left\{ \frac{1}{2\kappa}\left[R{+}2\Lambda {+} \lambda (g_{mn}U^m
	U^n {-}1 ){+} K^{abmn} \nabla_a U_m \nabla_b U_n \right] {+} \right.
	\label{1}
	\end{equation}
	$$
	\left. + \frac14 \left(F_{mn} + \phi  F^{*}_{mn}\right) F^{mn} + \frac{1}{2}\Psi^2_0 \left[V(\phi, \Theta, \sigma^2) {-} g^{mn}\nabla_m \phi \nabla_n \phi \right] \right\}
	\,.
	$$
	As usual, the term with the Lagrange multiplier $\lambda$ guaranties that the aether velocity four-vector is timelike and unit, $g_{mn}U^m U^n {=} 1$. The standard decomposition of the covariant derivative of the aether velocity four-vector
	\begin{equation}
	\nabla_iU_k = U_i DU_k + \sigma_{ik} + \omega_{ik}+ \frac13 \Theta \Delta_{ik} \,,
	\label{decompos}
	\end{equation}
	where $D=U^k \nabla_k$ is the convective derivative, $\Delta_{ik}=g_{ik}{-}U_iU_k$ is the projector, $\Theta = \nabla_k U^k$ is the expansion scalar, $\omega_{ik}=\frac12 \Delta^p_{i}\Delta^q_{k}(\nabla_pU_q {-} \nabla_qU_p)$ is the skew symmetric vorticity tensor, involves into consideration the  symmetric traceless shear tensor
	$\sigma_{ik}{=} \frac12\Delta^p_{i}\Delta^q_{k}\left(\nabla_pU_q {+} \nabla_qU_p)\right) {-}\frac13 \Theta \Delta_{ik}$ ($g^{mn}\sigma_{mn}{=}0$).
	Using the Jacobson's constitutive tensor
	\begin{equation}
	K^{abmn}{=} C_1 g^{ab} g^{mn} {+} C_2 g^{am} g^{bn}
	{+} C_3 g^{an} g^{bm} {+} C_4 U^{a} U^{b}g^{mn} \,,
	\label{K}
	\end{equation}
	one can rewrite the kinetic term  ${\cal K} \equiv K^{abmn}(\nabla_a U_m) (\nabla_b U_n)$  in the  following form
	\begin{equation}
	{\cal K} =(C_1 {+} C_4)DU_k DU^k {+}
	(C_1 {+} C_3)\sigma_{ik} \sigma^{ik} + (C_1 {-} C_3)\omega_{ik}
	\omega^{ik} {+} \frac13 \left(C_1 {+} 3C_2 {+}C_3 \right) \Theta^2
	\,. \label{act5n}
	\end{equation}
	Based on the results presented in \cite{170817} and on the theoretical predictions about the gravitational wave velocity in the aether \cite{J}, one has to put $C_3=-C_1$. This means that the shear tensor $\sigma_{ik}$ disappears from the decomposition (\ref{act5n}). The term in (\ref{1}), which contains the Maxwell tensor $F_{mn}$ and its dual $F^{*}_{mn}$ produces the equations of axion electrodynamics, when one uses the variation procedure with respect to the potential of the electromagnetic field $A_j$.
	The potential of the axion field $V$ is assumed to depend on the axion field $\phi$ itself, on the expansion scalar $\Theta$ and the square of the shear tensor $\sigma^2=\sigma_{pq} \sigma^{pq}$; the parameter $\Psi_0$ is reciprocal to the coupling constant of the axion-photon interaction.
	Due to the definition $F_{ik}=\nabla_i A_k {-} \nabla_k A_i$, we have to keep in mind the equation $\nabla_k F^{*ik} = 0$.
	The standard procedure of variation of (\ref{1}) with respect to $A_i$ gives the main equation of axion electrodynamics
	\begin{equation}
	\nabla_k \left[F^{ik} + \phi F^{*ik} \right] =0 \ \Rightarrow \ \nabla_k F^{ik} = -  F^{*ik} \nabla_k \phi \,.
	\label{2}
	\end{equation}
	Clearly, the properties of the magneto-electric configurations depend on the state of the axion field.

	Variation of (\ref{1}) with respect to $\phi$ gives the equation
	\begin{equation}
	g^{mn} \nabla_m \nabla_n \phi + \frac12 \frac{\partial }{\partial \phi}V(\phi, \Theta, \sigma^2) = - \frac{1}{4\Psi^2_0} F^*_{mn} F_{mn} \,.
	\label{ax10}
	\end{equation}
	The term in the right-hand side of this equation can be interpreted as the electromagnetic pseudoscalar source of the axion field; the second term in the left-hand side shows that the aether regulates the behavior of the axion field via the scalars $\Theta$ and $\sigma^2$.

	Variation with respect to $U^j$ gives us the equations of the aether dynamics in the canonic form
	\begin{equation}
	\nabla_a {\cal J}^{aj} = \lambda \ U^j  + I^j \,, \quad \lambda =  U_j \left[\nabla_a {\cal J}^{aj}- I^j \right]  \,,
	\label{0A1}
	\end{equation}
	but now the four-vector quantity $I^j$ and the tensor ${\cal J}^{aj}$ are, respectively, of the form
	\begin{equation}
	I^j =  C_4 (DU_m)(\nabla^j U^m) - \kappa \Psi_0^2 \frac {\partial V}{\partial (\sigma^2)} \sigma^{jm} DU_m \,,
	\label{0A22}
	\end{equation}
	\begin{equation}
	{\cal J}^{aj} = {\cal J}^{(0)aj}  + \kappa \Psi_0^2 \left [\frac {\partial V}{\partial (\sigma^2)} \sigma^{aj}+\frac12 \frac {\partial V}{\partial \Theta} g^{aj}\right] \,, \quad
	{\cal J}^{(0)aj} = {K}^{abjn} (\nabla_b U_n) \,.
	\label{0A33}
	\end{equation}
	Equations for the gravitational field as the result of variation with respect to metric is of the form
	\begin{equation}
	R_{ik} - \frac{1}{2} R \ g_{ik}
	=  \Lambda g_{ik}  + \kappa T^{(EM)}_{ik} + \kappa T^{({\rm A})}_{ik} +  T_{ik}^{(U)} + T_{ik}^{(V)}  \,. \label{0Ein1}
	\end{equation}
	The stress-energy tensors of the electromagnetic and pure axion field have the standard form
	\begin{equation}
	T^{({EM})}_{ik} = \frac14 g_{ik} F_{mn} F^{mn} {-} F_{im} F_k^{\ m}\,, \quad T^{(A)}_{ik} = \Psi^2_0 \left[\nabla_i \phi \nabla_k \phi
	+\frac12 g_{ik}\left(V {-} \nabla_n \phi \nabla^n \phi \right) \right] \,.
	\label{qq1}
	\end{equation}
	Other contributions contain the derivatives of the axion field potential with respect to $\Theta$ and $\sigma^2$:
	$$
	T_{ik}^{(U)} =
	\frac12 g_{ik} \ K^{abmn} \nabla_a U_m \nabla_b U_n{+} U_iU_k U_j \nabla_a {\cal J}^{aj} {+} C_4 \left(D U_i D U_k {-} U_iU_k DU_m DU^m \right) {+}
	$$
	\begin{equation}
	{+}\nabla^m \left[U_{(i}{\cal J}^{(0)}_{k)m} {-}
	{\cal J}^{(0)}_{m(i}U_{k)} {-}
	{\cal J}^{(0)}_{(ik)} U_m\right]{+}
	C_1\left[(\nabla_mU_i)(\nabla^m U_k) {-} (\nabla_i U_m )(\nabla_k U^m) \right]
	\,,
	\label{5Ein1}
	\end{equation}
	\begin{equation}
	T_{ik}^{(V)} = \kappa \Psi_0^2 \left \{2 \frac {\partial V}{\partial (\sigma^2)} \left[\sigma_{m(i}\sigma_{k)}^{m} {-} \sigma_{m(i}\nabla_{k)}U^m {+} \frac13 \Theta \sigma_{ik}\right] {-} (D{+}\Theta)\left [\sigma_{ik} \frac {\partial V}{\partial (\sigma^2)}  {+} \frac12 g_{ik} \frac {\partial V}{\partial \Theta} \right]\right \} \,.
	\end{equation}
	
	\section{Bianchi-I spacetime platform}
	
	When one considers the anisotropic spatially homogeneous Universe with the metric
	\begin{equation}
	ds^2 = dt^2 - a^2(t)dx^2 - b^2(t)dy^2 - c^2(t)dz^2 \,,
	\label{App1}
	\end{equation}
	and with the aether velocity four-vector $U^j = \delta^j_0$, one obtains immediately that
	\begin{equation}
	\nabla_i U_k = \nabla_k U_i  = \frac12 \dot{g}_{ik} \Rightarrow \ DU^i = 0 \,, \quad \omega_{mn}=0 \,, \quad \Theta = \frac{\dot{a}}{a} + \frac{\dot{b}}{b}+ \frac{\dot{c}}{c} \,, \quad \sigma_{0k} = 0 \,,
	\label{App2}
	\end{equation}
	\begin{equation}
	\sigma^1_1 = \frac{\dot{a}}{a} {-} \frac13 \Theta \,, \quad \sigma^2_2 = \frac{\dot{b}}{b} {-} \frac13 \Theta \,, \quad \sigma^3_3 = \frac{\dot{c}}{c} {-} \frac13 \Theta \,, \quad
	\sigma^2 = \left(\frac{\dot{a}}{a} \right)^2 {+} \left(\frac{\dot{b}}{b} \right)^2 {+}\left(\frac{\dot{c}}{c} \right)^2 {-} \frac13 \Theta^2 \,.
	\label{App3}
	\end{equation}
	Next, we will have a thorough analysis of the complete system of model equations taking into account the term $\sigma^2$ (\ref{App3}).
	
	\section*{Conclusion}
	
	The full-format analysis of the system of equations for gravitational, electromagnetic, vector and pseudoscalar (axion) fields is, unfortunately, out of the frames of this short note. However, based on the presented results, we can see  that the incorporation of the square of the shear tensor $\sigma^2$ into the axion field potential $V(\phi, \Theta, \sigma^2)$ removes the degeneration with respect to the shear tensor, attributed to the aether velocity.


\end{document}